# Reversing strain deformations probe mechanisms for enhanced segmental mobility of polymer glasses

Kelly Hebert and M.D. Ediger*

Department of Chemistry, University of Wisconsin-Madison, Madison, WI, 53706, United States

*To whom correspondence should be addressed; ediger@chem.wisc.edu

## Abstract

Optical probe reorientation measurements were performed to monitor changes in segmental dynamics resulting from the nonlinear deformation of a polymer glass. Segmental dynamics were monitored in a poly(methyl methacrylate) glass near $T_g$ before and after a series of reversing deformations in which the sample was extended at constant strain rate and then allowed to retract back to zero stress at constant strain rate. Evidence of a rejuvenation mechanism, as quantified by a departure of the segmental dynamics from the quiescent aging dynamics after the reversing deformation, is observed for deformations which reach 60% of the yield strain or greater. By this measure, a saturation of the rejuvenation mechanism is not observed until at least five times the yield strain. For comparison, purely mechanical measurements of rejuvenation, based upon the reduction of the yield stress in a subsequent deformation, were also performed. These purely mechanical experiments show broad qualitative agreement with the probe reorientation experiments, but quantitatively differ in the pre-yield regime. The results are discussed in the context of recent theoretical approaches and simulations which provide a molecular-level description of polymer glass deformation.

**Introduction**

In spite of significant modeling, simulation, and experimental effort, a fundamental understanding of the deformation behavior of polymer glasses is still being developed. Among the earliest work was that of Eyring,[1] who concluded that stress acts to lower barriers for rearrangements during the deformation of a solid. Since then, many other models have been developed to understand the deformation behavior of polymer glasses.[2-13] Integral to the work of these later workers is the idea that during nonlinear deformation, the segmental dynamics (local rearrangements involving a few repeat units along the chain)[14] of a polymer glass becomes significantly faster, which then allows flow to occur at much lower stress than would be predicted by linear viscoelasticity. This interpretation is supported by recent simulations that have observed enhanced segmental dynamics during constant strain rate[15-18] and constant stress[16, 19, 20] deformation. Experimentally, changes in segmental dynamics during deformation have been measured by NMR,[21] diffusion,[22] dielectric spectroscopy,[23, 24] and probe reorientation.[19, 20, 25-31] The probe reorientation measurements show that the observed decrease in the average segmental relaxation time during deformation (up to 3 orders of magnitude) can roughly account for the observed flow stress,[32] supporting the view that enhanced segmental dynamics is the key to understanding the nonlinearity of polymer glass deformation.[32]

An aspect of the deformation of polymer glasses which is not captured by the simple Eyring picture is the influence of thermal and mechanical history on deformation. For constant strain rate experiments, increasing the aging time prior to deformation generally results in an increase in the yield stress.[2] In contrast, pre-deforming a glass sample beyond yield reduces the yield stress.[33-36] Since pre-deformation apparently acts to undo the effects of aging, this process has been described as mechanical "rejuvenation".[37]

A simple interpretation of mechanical rejuvenation and the Eyring mechanism can be given in terms of the potential energy landscape (PEL).[38] Physical aging acts to equilibrate a polymer glass and in the process the position of the system in the PEL is lowered. The activation barriers for rearrangements are expected to increase in lower regions of the PEL; this is consistent with the super-Arrhenius temperature dependence of polymer melts in equilibrium. The higher activation barriers that result from aging will naturally lead to a higher yield stress. Since mechanical rejuvenation reverses the effects of aging, at least to a first approximation, we infer that rejuvenation pulls the system into higher regions of the PEL where the activation energies are smaller, leading to a lower yield stress. The Eyring mechanism can also be interpreted in terms of the PEL as landscape tilting: the landscape tilting mechanism lowers barriers for those rearrangements that are favored by the applied force, but those rearrangements do not necessarily pull the system higher in the PEL.

There are several models for polymer glass deformation[2, 8-12] that combine an Eyring-like stress activation mechanism with mechanical rejuvenation. Qualitatively, these models can account for the effects of stress, physical aging, and pre-deformation on the mechanical properties of polymer glasses. One theory incorporating the landscape tilting and rejuvenation mechanisms, the Nonlinear Langevin Equation (NLE) theory of Chen and Schweizer,[8, 9] predicts changes in segmental dynamics during constant stress and constant strain rate deformation consistent with simulations[15, 39] and probe reorientation experiments.[26, 30, 31] Calculations using the theory to investigate constant strain rate deformation of poly(methyl methacrylate) glasses indicate that up until ~60% of the yield strain, landscape tilting is the sole mechanism acting to enhance segmental dynamics.[9] At higher strains, the rejuvenation mechanism also becomes active. In the theory, the rejuvenation mechanism is physically tied to an increase in $S_0$, the amplitude of nanometer-scale density fluctuations in the glass. As strain increases beyond 60% of the yield strain, $S_0$ is predicted to slowly increase and then to saturate at strains significantly past yield; the theory predicts that rejuvenation is not complete until the strain reaches at least five times the yield strain.

There has been significant controversy about when rejuvenation begins during deformation and even disagreements about the defining features of rejuvenation. The mechanical deformation of polymer glasses was extensively explored by Struik.[40] In one type of experiment, stresses below the yield-stress were observed to raise the creep compliance to values characteristic of a less-aged glass. Struik interpreted this and the related "stress pulse" experiment to indicate that pre-yield deformation can at least partially erase the effects of prior aging. However, this interpretation has been challenged by McKenna[35] on the basis of experiments which simultaneously measured mechanical deformation and sample volume. Santore, Duran, and McKenna observed that while pre-yield torsional deformation of an aging polymer glass transiently increased the sample volume, the sample volume rejoined the original aging curve such that the time to reach equilibrium was not extended.[41] For Santore et al., the observed time dependence is inconsistent with mechanical rejuvenation while Struik later stated just the opposite.[42] Hasan and Boyce[43] measured the enthalpy of deformed polymer glasses and observed only very small enthalpy changes in pre-yield deformations, consistent with the idea that pre-yield deformation has little effect on the position of the glass on the PEL (i.e., pre-yield deformation does not rejuvenate the glass).

Here we present experiments in which the segmental dynamics of a polymer glass are monitored following reversing constant strain rate deformations. Our goal is not to resolve all the controversies surrounding mechanical rejuvenation, but rather to allow a targeted comparison with recent theoretical work that explains changes in segmental dynamics as resulting from the landscape tilting and rejuvenation mechanisms. While many different responses can be used to judge the extent to which a particular deformation has induced rejuvenation, we focus on the segmental dynamics due to its central role in the nonlinearity of polymer glass deformation. We performed a series of constant strain rate deformations on a poly(methyl methacrylate) glass at $T_g$ – 7 K to varying strains which were then immediately reversed at the same absolute strain rate until a near-zero stress is achieved. Using an optical probe reorientation technique, changes in segmental dynamics are tracked before and after the reversing deformations. For models that explain changes in segmental dynamics as resulting from landscape tilting and rejuvenation, the zero-stress condition that is maintained after the deformation means that any subsequent enhancements of segmental dynamics should be ascribed to rejuvenation – in this paper, unless otherwise qualified, "rejuvenation" has this specific meaning. In addition to our probe reorientation measurements, we performed purely mechanical experiments in which a second constant strain rate deformation follows the reversing deformation; the decrease in the yield stress observed in the second deformation is used to provide a purely mechanical perspective on rejuvenation that is similar to previous investigations.[33-36] In contrast to the probe reorientation measurements which do not perturb the glass properties, measurements of decreased yield stress is a major perturbation and this must be considered in their interpretation. We compare our optical and mechanical rejuvenation results to those calculated based upon the increase of $S_0$, the amplitude of nm-scale density fluctuations, in the NLE theory.[9] Although calculations from the NLE theory[8, 9] are used for comparison in this study, the present experimental work provides a general characterization of the nature of rejuvenation mechanisms that can be used to test other theoretical and modeling work.

We find that, after reversing deformations, both the optical and mechanical experiments indicate that rejuvenation develops gradually with strain and does not saturate until at least five times the yield strain. In this respect, our results agree with published calculations of the increase of $S_0$ in the NLE theory.[9] We also find that substantial levels of rejuvenation are observed via optical measurements after reversing deformations in the pre-yield regime. These pre-yield rejuvenation effects observed through changes in segmental dynamics are broadly consistent with simulations of Smessaert and Rottler.[44] In contrast, the mechanical experiments of this work and calculations from the NLE theory[9] demonstrate little rejuvenation activity in the pre-yield regime. We discuss our results in light of previous experiments and recent simulations and theory.

## Experimental Methods

### *Sample Preparation*

Lightly cross-linked poly(methyl methacrylate) (PMMA) samples were synthesized following a previously-reported procedure.[25, 30] A solution containing 1.5 weight percent ethylene glycol dimethacrylate (EGDMA, Polysciences, Inc.) in methyl methacrylate (MMA, Polysciences, Inc.) was used to create the cross-linked films. The fluorescent probe N, N'-Dipentyl-3,4,9,10-perylenedicarboximide (DPPC, Aldrich) was also dissolved in this stock solution at a concentration of ~$10^{-6}$ M . The initiator benzoyl peroxide (Polysciences, Inc.) was added to this stock solution at a concentration of 0.1 weight percent, and the resulting mixture was heated at 343 K for approximately 30 minutes. This thickened mixture was subsequently loaded into molds consisting of two microscope slides lined with aluminum foil spacers; these filled molds were then clamped together. The stock solution was then allowed to polymerize in the molds in a 363 K oven for 24 hours under a positive pressure of nitrogen gas. The oven temperature was then increased to 413 K for an additional 24 hours. Films were removed from their molds through sonication in a water bath. Individual samples were cut from the resulting films by a die cutter. The shape of the die cutter is a 50% miniaturization of the dimensions prescribed by ASTM method D1708-10.[45] The polymerization procedure results in films that typically range from ~35-50 μm at the edges and ~25-30 μm in the middle of the sample. The samples had a glass transition temperature $T_g$ of 395 ± 1 K, determined using differential scanning calorimetry (DSC) from the onset during the second heating scan at a rate of 10 K/min. Using conditions similar to that described above (but without the EGDMA crosslinking agent), Caruthers and coworkers synthesized uncrosslinked PMMA of $M_w$=2.24 × $10^6$ g/mol, as determined by GPC.[46] In light of these high molecular weights, free chain ends are not anticipated to significantly influence the reported dynamics for the cross-linked material utilized in the current work.

For the results reported in this manuscript, a single PMMA sample was used for all the optical measurements of rejuvenation and two samples were used for the mechanical measures of rejuvenation; however, additional tests performed on other samples yielded results consistent with the reported results. Between subsequent optical and mechanical experiments, the sample was allowed to anneal above $T_g$ (408 K) for at least three hours. This annealing step resulted in the full erasure of imposed strain from prior tests. As a further check, we ran experiments out of order with respect to the maximum strain achieved during deformation and found good overlap in both the mechanical and pre-deformation optical data, as shown in Figures 2 and 3.

*Deformation Instrumentation*

Samples were held in a previously-described deformation apparatus[30] for the entirety of each experiment, including annealing above the glass transition temperature. A programmable linear actuator is attached to a u-shaped bar which is then coupled to a load cell and the sample. Increasing the position of the actuator drives tensile deformation, while decreasing the position of the linear actuator allows previously-imposed strain to decrease when the sample is under tension.

Stress was calculated based on the initial cross-section of the thinnest portion of the sample, which was 2.0 mm × 25-30 μm, and the force reported by the load cell; global strain was calculated based on movement of the linear actuator. We perform probe reorientation measurements in a ~500 μm × ~500 μm measurement area at the thinnest region of the sample. In addition to monitoring the global strain, we monitor strain in this local measurement area by photobleaching lines into the sample before deformation is applied. We can then calculate local strain by imaging these lines. Our samples are not uniform in thickness, and thus do not deform homogeneously after yielding occurs. Below the yield strain of 0.029, the global and local strains are approximately equal. In the post-yield regime, strain builds up fastest in the thinnest area of the sample where we perform our optical experiments; the local strain rate is approximately 2-3 times the global strain rate after yielding occurs. We ensured that we were performing the optical measurements in the thinnest portion of the sample by pre-deforming a sample and photobleaching a mark at the location where a neck-like region first formed after yield, which could be easily identified through a wide-field view under our microscope. The sample was then annealed above $T_g$ for at least three hours, resulting in the full erasure of strain in the sample, and the bleached location could be visually identified for subsequent experiments. We found that the relationship between the local strain and global strain during deformation was reproducible for experiments within one sample. This was confirmed by capturing several images of the measurement area during the tensile deformation phase of multiple experiments. The local strain vs. global strain data for multiple tests collapsed onto a single curve, which was then fitted to a $7^{th}$ order polynomial[30] to determine the maximum local strains attained during each experiment.

*Thermal Protocol*

Samples were loaded into and held within a temperature-controlled cell during all phases of the experiment, including annealing above $T_g$, cooling to the testing temperature, and during aging and deformation. To erase thermomechanical history, samples were annealed at 408 K for at least three hours and then cooled at 1 K/min to the testing temperature of 388 K. After cooling, the temperature remains constant to 0.2 K. Reported temperatures are accurate to ± 1 K as determined by melting point tests performed within the brass cell. Times reported in this study reflect time spent below the glass transition temperature as determined by DSC (395 ± 1 K). A heating rate of 10 K/min rather than 1 K/min was used for $T_g$ determination; the glass transition temperature of our samples at the imposed cooling rate is likely ~3 K lower, resulting in a slight overestimation (~180 s) in the reported aging times in this study; as this is very small compared to the aging time, this should have no measurable effect on the reported results.

*Mechanical Protocol*

A schematic of the mechanical protocol used for the probe reorientation experiment is shown in Figure 1a. Prior to deformation, the sample was held without imposed stress while it aged at 388 K. After a predetermined aging time (9300 s), a global engineering strain rate of $1.55 \times 10^{-4}$ $s^{-1}$ was imposed on the sample up to a pre-determined strain, after which strain was immediately allowed to reverse at the same strain rate. This unloading continued until a stress of 0.2 MPa was reached, at which time the linear actuator was programmed to retract in order to maintain $\leq$ 0.2 MPa stress for the remainder of the experiment. This small imposed stress of 0.2 MPa after the reversing deformation was used to stabilize the position of the sample for further optical experiments; we do not expect this stress to have a measurable impact on reported segmental dynamics. To test this, we repeated this procedure while instead imposing a 1 MPa stress after the deformation and found no significant difference in the optical data. We emphasize that strain reversal in this work indicates unloading to a near-zero stress, rather than applying a compressive deformation to achieve zero strain.

In addition to the optical measurements of probe mobility, we employ a purely mechanical protocol to quantify rejuvenation as shown in Figure 1b. Changes in the post-deformation mechanical behavior of a polymer glass have been previously used to investigate the effect of mechanical rejuvenation in a glass; for example, Govaert and coworkers have investigated the reduction and subsequent regrowth of the yield stress after large mechanical deformations.[33, 34] The mechanical protocol shown in Figure 1b is identical to that shown in Figure 1a, except that after a waiting period of 1200 seconds after the reversing deformation, a second deformation at a constant global strain rate of $1.55 \times 10^{-4}$ $s^{-1}$ is performed. While the mechanical procedure that we use to quantify rejuvenation is not guided by a theory or model, we regard it as a reasonable choice for comparison with the optical measurements given previous observations of reduced yield stress after large deformations as reported by, for example, Govaert,[33, 34] McKenna,[35] Bauwens,[36] and their coworkers.

Although large pre-deformations typically result in the full erasure of the overshoot peak in a constant strain rate deformation, we observe a small overshoot in the mechanical data after the largest reversing deformation as can be seen below in Figure 4a. We attribute this remaining overshoot to physical aging during the imposed 1200 second waiting time between the end of the reversing deformation and the second constant strain rate deformation. Measurable regrowth of the overshoot peak was observed almost immediately after a large deformation in the work of Govaert and coworkers,[34] which investigated polystyrene glasses held at room temperature. We have performed additional mechanical testing in which no waiting time was imposed between the reversing deformation and subsequent constant strain rate deformation and found essentially complete erasure of the overshoot peak. The waiting time in our protocol is imposed to provide a fair comparison to probe reorientation data such that both methods reflect the same waiting period after the reversing deformation; further discussion of this point will be provided below.

A temperature increase has been previously reported to accompany the yielding of polymer glasses,[47] such as the glassy PMMA studied in the present work. For the deformations studied in the present work, any such temperature increase should be negligible. If we make the extreme assumption that no heat is dissipated during deformation and that all mechanical work directly acts to heat the sample, the maximum increase in temperature during these experiments would be 1.2 K. However, a study of Haward and coworkers[47] found that temperature changes during the deformation of a glassy polymer are strongly dependent on the strain rate, with smaller increases in temperature experienced for lower strain rates. For strain rates of 0.016 mm/s (~3 times faster than the strain rates used in the present study), Haward and coworkers found that less than 20% of the applied mechanical work converted to measurable heating of their samples. Additionally, the samples used in the present study are more than 35 times thinner than the thinnest samples used in the study of Haward and coworkers; any heat dissipation during deformation should be accordingly faster.

*Probe Reorientation Protocol*

Segmental dynamics before and after the reversing deformation were monitored through a previously-described probe reorientation technique.[19, 25-31] Reorientation of an anisotropic ensemble of fluorescent DPPC probes tracks segmental dynamics in the PMMA matrix during aging and deformation; the anisotropy is induced by photobleaching with a linearly-polarized 532 nm laser beam. Probe reorientation is tracked by monitoring fluorescence in the remaining unbleached population of probe molecules in response to a weak, circularly-polarized 532 nm laser beam. Fluorescence intensities in the polarization directions parallel and perpendicular to the original photobleaching polarization are monitored. The anisotropy of the ensemble of fluorescing probes can then be calculated from these intensities as described previously.[25] The time-dependent anisotropy decay can be described by the Kohlrausch-Williams-Watts (KWW) function:

$$r(t) = r(0) \cdot e^{-(t/\tau_{1/e})^{\beta_{KWW}}}$$

Here r(t) and r(0) are the time-dependent and initial anisotropies, respectively. Fitting the time-dependent anisotropy to the KWW function yields the time for the anisotropy to decay by a factor of e, $\tau_{1/e}$, and the stretching exponent $\beta_{KWW}$. The reorientation of the DPPC probes has been previously reported to be a good reporter for the α (segmental) relaxation of PMMA above $T_g$ in the absence of deformation as evidenced by probe reorientation and $\tau_\alpha$ from dielectric spectroscopy displaying the same temperature dependence under these conditions.[27] We make the assumption that the probes are also good reporters below $T_g$ and during deformation, which is consistent with the following observations. Changes in segmental dynamics as reported by probes during constant stress[19, 20, 25-29] and constant strain rate deformation[30, 31, 48] demonstrate behavior that is qualitatively consistent with those observed in computer simulations (which do not incorporate probes).[15, 19, 20, 39] Additionally, preliminary experiments demonstrate that during physical aging of a polymer glass close to $T_g$, probe reorientation tracks mechanical relaxation times as determined by stress relaxation in the linear response regime.

For the experiments reported here, in which segmental dynamics were not monitored during active deformation, $\beta_{KWW}$ did not significantly change following deformation and was fixed to a value of 0.31 in our fitting procedure. Data was also analyzed without fixing $\beta_{KWW}$; although the data showed more noise, the general trends in the data did not change. Although previous work has shown that $\beta_{KWW}$ can change during deformation,[20, 26, 27, 31] considering the temperature, strain rate, and reversing protocol used here,[29] our finding of an essentially constant $\beta_{KWW}$ is consistent with previous work.

**Results**

In this section, we present results describing how reversing constant strain rate deformations change the subsequent segmental dynamics and mechanical properties of a polymer glass. Implications of the data in light of existing theories and simulations will be outlined in the Discussion section.

Figure 1 shows a schematic of the protocols that we used. Figure 1a shows the probe reorientation measurement protocol; Figure 1b shows the purely mechanical protocol. In each case, lightly cross-linked PMMA samples were held below $T_g$ in the absence of stress or strain for 9300 seconds, at which time a reversing constant strain rate deformation was applied. For each deformation, a rate of $1.55 \times 10^{-4}$ $s^{-1}$ was applied to a set strain and then the strain was allowed to reverse at the same rate until a stress of 0.2 MPa was reached. The total time elapsed during the reversing deformation ranged from 110 to 1020 seconds for these experiments. Further retraction of the sample was then programmed to maintain a stress of 0.2 MPa or less. For the experiments shown in Figure 1a, optical measurements of probe reorientation were performed both before and after the reversing deformation. The protocol for the purely mechanical experiment shown in Figure 1b is identical to Figure 1a with the exception of an additional constant strain rate deformation performed 1200 seconds after the reversing deformation. The time at which the rejuvenation mechanism is evaluated in the probe reorientation measurement is represented by the star in Figure 1a, which also corresponds to the beginning of the second constant strain rate deformation in Figure 1b. Further discussion of the post-deformation stress and imposed waiting time may be found in the Experimental Methods section. The time axis in Figure 1 is not shown to scale in order to more clearly highlight the deformation protocol.

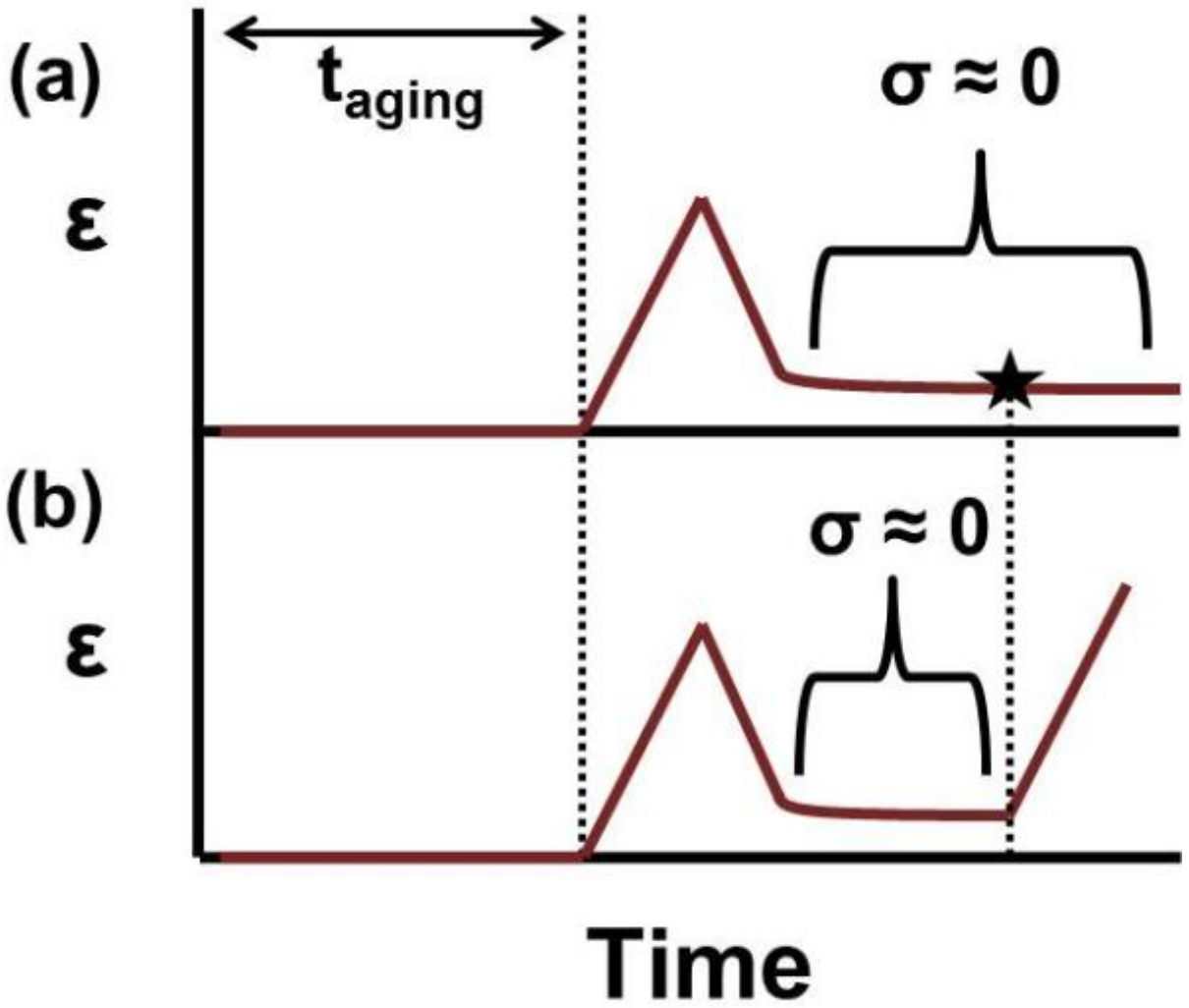

**Figure 1.** Schematic diagram illustrating the mechanical protocol utilized during probe reorientation experiments (a) and purely mechanical experiments (b). In panel (a), $\tau_{1/e}$ for the purposes of assessing the rejuvenation mechanism was reported for a fixed time after the end of the deformation, as shown by the star. This same fixed time was imposed as a waiting time between the reversing deformation and the second constant strain rate deformation in panel (b).

Figure 2 shows mechanical data collected during the deformation phase of Figure 1a. Reversing deformations spanning the pre-yield, yield, and strain softening regimes are represented. Probe reorientation measurements were monitored in a local region of the sample and strains reported in the legend represent the maximum strain attained in the local measurement region. The time axis in Figure 2 represents the total elapsed time since the sample reached the DSC $T_g$ of 395 K upon cooling; deformation commences 9300 seconds after $T_g$ is reached.

Figure 3 shows the evolution of the probe reorientation time ($\tau_{1/e}$) before and after the reversing deformations shown in Figure 2. In the absence of deformation, $\tau_{1/e}$ evolves toward equilibrium as shown by the solid line, which represents a power law fit to the aging data: $\tau_{1/e} \propto$

$t^{+0.72}$. After small pre-yield reversing deformations (e.g., see blue squares), very little change is seen from this base aging trajectory; however, a significant departure from the aging trajectory is seen in the case of large pre-yield deformation. As can be seen in Figure 3, the decrease of $\tau_{1/e}$ from the aging trajectory appears to saturate after deformations which reach several times the yield strain. It should be emphasized that all data points in Figure 3 were collected at a stress of essentially zero. The actual stress was 0.2 MPa or less, and this does not impact the measured $\tau_{1/e}$ as discussed in the Experimental Methods section.

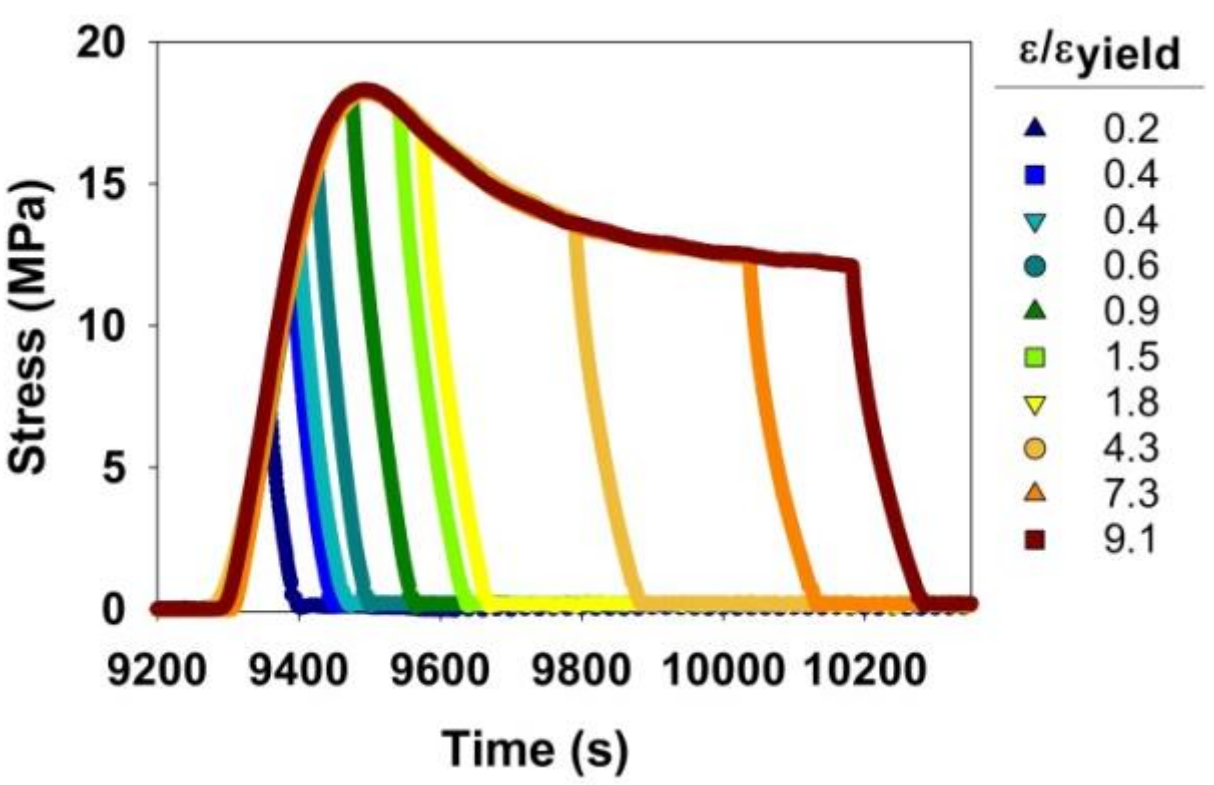

**Figure 2.** Evolution of stress during the reversing deformation for the protocol outlined in Figure 1a. All deformations were started at a fixed aging time of 9300 seconds and reversed at different strains. Strains reported in the legend reflect the maximum strain attained in the local measurement area during the deformation, relative to the yield strain. At the conclusion of the deformation, a stress of 0.2 MPa or less was maintained for the remainder of the experiment.

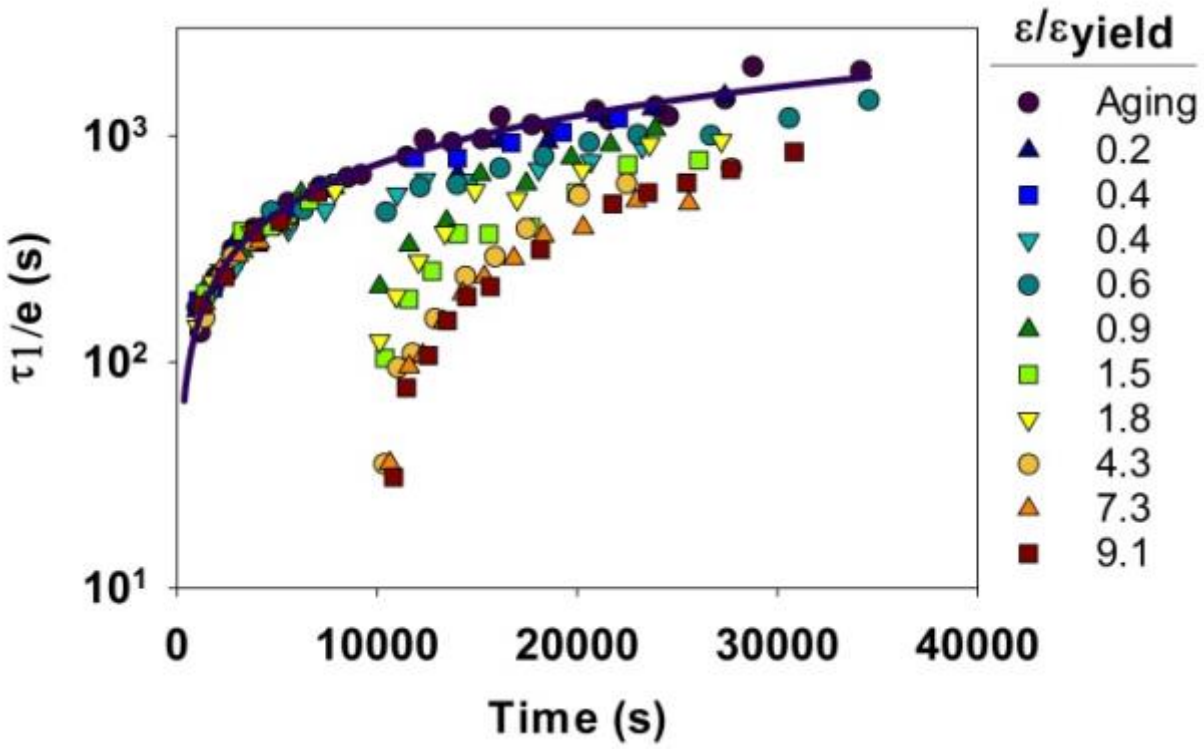

**Figure 3.** Evolution of the probe reorientation time $\tau_{1/e}$ before and after the reversing deformation. Deformation was started after aging for 9300 seconds and reversed at different strains. Strains reported in the legend reflect the maximum strain attained in the local measurement area during the deformation, relative to the yield strain. The solid line is a fit to physical aging data obtained in the absence of deformation and represents $\tau_{1/e} \propto t^{+0.72}$.

Using the probe reorientation data in Figure 3, we can determine the extent to which the sample has been rejuvenated, at least within the framework of models that include the landscape tilting and mechanical rejuvenation mechanism. In this framework, the zero-stress condition at the end of the experiment means that any observed enhancement of segmental dynamics should be ascribed to rejuvenation. The level of rejuvenation at a fixed time after the reversing constant strain rate deformation was determined by the decrease in log($\tau_{1/e}$) relative to the aging curve. Because all of the reversing deformations began at the same aging time but extended to various strains, the end time of each reversing deformation increased with increasing strain; for each deformation, $\tau_{1/e}$ was determined at 1200 seconds after the end of deformation by interpolation of the data (or by slight extrapolation, in one case). Rejuvenation was then determined by setting a linear scale from 0% - 100% based on the decrease in log($\tau_{1/e}$) from the aging trajectory, where 0% represents no change from the aging curve and 100% rejuvenation represents the ~1.3 decade decrease in $\tau_{1/e}$ found after the deformation of the largest strain; these results are presented below in Figure 5.

We also utilized a purely mechanical method to compare to the probe reorientation results presented in Figure 3. As described in the introduction, the extent to which the stress overshoot is reduced by pre-deformation is often used as a qualitative indication of the extent of rejuvenation.[33, 34] Figure 4 shows the results from two experiments of this type. After each reversing constant strain rate deformation, the glass was held for 1200 seconds at ~0.2 MPa before a second constant strain rate deformation was applied as outlined in Figure 1b. Figure 4a shows the stress measured in an initial deformation to a strain of 5.6 $\varepsilon_{yield}$, the subsequent decrease of the stress as the strain is reduced, and after an imposed period of near-zero strain, the stress measured during a second constant strain rate extension. Of interest is the near-erasure of the large overshoot peak in the second deformation; we attribute the small overshoot remaining to partial recovery during our imposed 1200 second waiting time (see Experimental Methods section). For comparison, Figure 4b shows analogous data for which the reversing strain reached a maximum value of 0.8 $\varepsilon_{yield}$. In contrast to panel a, the yield stress in the second deformation is only slightly reduced, even though the first deformation nearly reached the yield strain. In quantifying data like that shown in Figures 4a and 4b, we use the decrease of the yield stress (difference between the black and green dashed lines) to assess rejuvenation. Similar to the probe reorientation measurement, rejuvenation was evaluated on a linear scale of 0% to 100% based on the reduction of the yield stress, where 0% rejuvenation corresponded to the yield stress in the absence of prior deformation (16.9 MPa) and 100% rejuvenation corresponded to the yield stress for a reversing deformation out to large strains (11.7 MPa). The limitations in using the yield stress as an indicator of rejuvenation will be considered in the Discussion section.

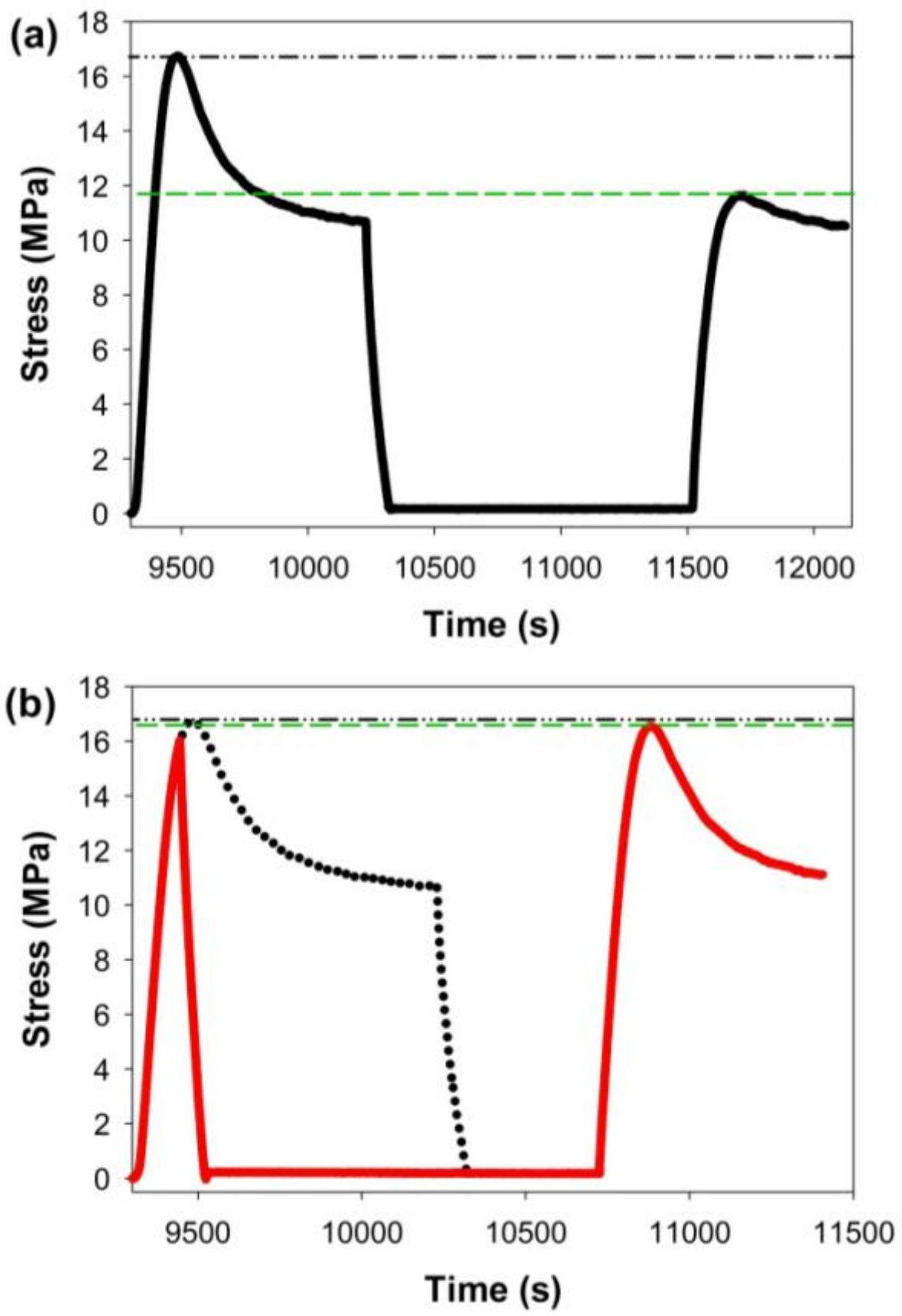

**Figure 4.** Two representative sets of mechanical data obtained using the protocol described in Figure 1b. Panel (a) shows the largest post-yield deformation (reversed at 5.6 $\varepsilon_{yield}$), resulting in a significant reduction in the yield stress upon a second constant strain rate deformation. Panel (b) demonstrates that a reversing deformation to 0.8 $\varepsilon_{yield}$ (red) does not significantly alter the subsequent mechanical properties of the glass. In each panel, the black and green dashed lines mark the original and post-deformation yield stresses, respectively. The reversing portion of the mechanical data in Figure 4a is overlaid in Figure 4b (black dots) as a reference.

Figure 5 compares the extent of rejuvenation after reversing deformations for the probe reorientation experiment and the purely mechanical experiment. At strains greater than the yield strain (solid vertical line), the mechanical and optical measures of rejuvenation show good agreement, with the extent of rejuvenation gradually increasing until roughly five times the yield strain. For strains lower than the yield strain, the probe reorientation method displays much greater evidence of rejuvenation as compared to the mechanical experiments. The solid line shows fractional rejuvenation based on theoretical calculations from reference 9; these will be discussed further below.

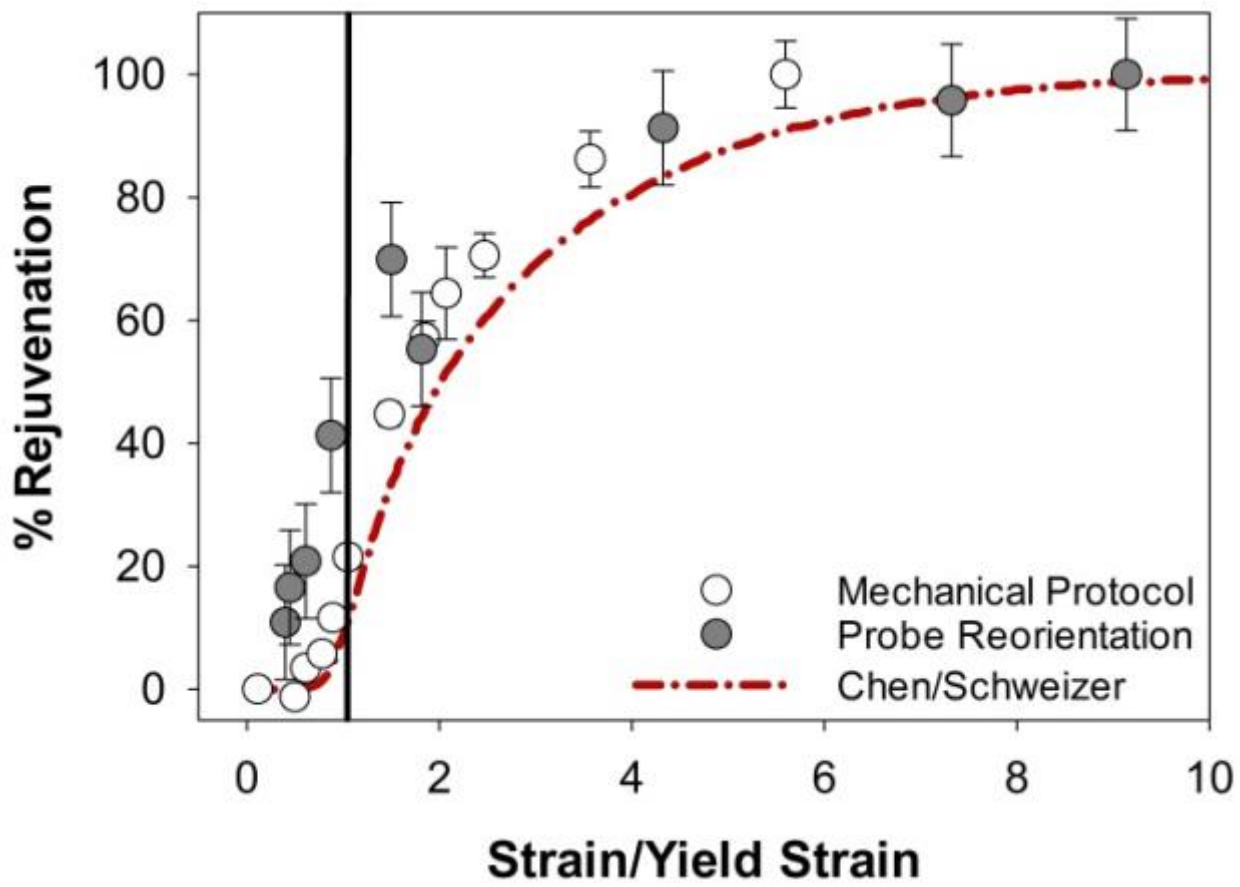

**Figure 5.** Fractional rejuvenation vs. normalized strain as calculated through the probe reorientation measurement (reduction in $\tau_{1/e}$), the purely mechanical protocol (reduction in yield stress), and from reference 9 (increase in $S_0$ during constant strain rate deformation). Data points represent measurements conducted 1200 seconds after the conclusion of the reversing deformation. The dashed line represents calculations from reference 9 for a PMMA glass deformed in tension at a constant strain rate (see text for details). The solid line marks strain at yield. For the x-axis, local and global strains are used for probe reorientation and purely mechanical experiments, respectively.

## Discussion

Figures 3-5 demonstrate a gradual increase in the extent of rejuvenation, as deduced by various methods, with increasing strain during reversing constant strain rate deformations. These figures also demonstrate a difference in the extent of rejuvenation as measured by probe

reorientation and the purely mechanical stress overshoot experiment. In this Discussion section, we first compare the optical probe results to recent theoretical and modeling work. We then compare the recovery of segmental dynamics after reversing deformations to simulations of Smessaert and Rottler.[44] Finally, we explore the difference between the optical and mechanical measures of rejuvenation, and compare our results with previous experimental work.

*Comparison of probe reorientation results to theoretical and modeling work*

In their NLE theory, Chen and Schweizer[8, 9] include two mechanisms that allow for enhancement of segmental dynamics during polymer glass deformation. The first of these mechanisms is what we describe as the landscape tilting mechanism, in which stress acts to lower free energy barriers for rearrangements, but does not change the position of the system on the potential energy landscape. The second of these mechanisms is the rejuvenation mechanism, which takes effect at sufficiently high strains and can act concurrently with the landscape tilting mechanism. In the theory, physical aging decreases the amplitude of nm-scale density fluctuations ($S_0$) while deformation increases $S_0$. We interpret this to mean that deformation can drive the system higher on the potential energy landscape where barriers are lower. Using the NLE approach, Chen and Schweizer[8, 9] predict changes in segmental dynamics during constant stress and constant strain rate deformations that are consistent with reported experimental[26, 30, 31] and simulation[15, 39] studies. In these previous experiments, landscape tilting and rejuvenation can act together to enhance segmental dynamics. During constant strain rate experiments, for example, the NLE theory predicts that both mechanisms are active in enhancing dynamics in the post-yield regime and that the relative importance of the two mechanisms changes with strain even while the segmental relaxation time is constant.[9] While the experiments do indicate a constant segmental relaxation time in the post-yield regime,[9] experiments published up to the present have not separately tested these two mechanisms.

One goal of the present work is to separately test the rejuvenation mechanism in the NLE approach. In the framework of this theory, the landscape tilting mechanism is only active when stress is applied to the glass. Thus, after the reversing deformations in the present study, any enhancement of dynamics due to the landscape tilting mechanism is removed, and any enhanced dynamics can be attributed to the rejuvenation mechanism. By eliminating the landscape tilting mechanism contributions from our post-deformation measurements of $\tau_{1/e}$, we determine the level of strain required for the rejuvenation mechanism to become active.

Figure 5 shows the activity of the rejuvenation mechanism in the NLE theory (dot-dash line),[8, 9] as characterized by $S_0$, as a function of strain *during tensile constant strain rate deformation*. These previously-published calculations for $S_0$ are presented on a fractional rejuvenation scale, where the rejuvenation level of 0% corresponds to the value of $S_0$ at the start of deformation, and 100% rejuvenation corresponds to the value of $S_0$ in the limit of large strain. At small strains, $S_0$ remains constant (indicating that the rejuvenation mechanism is not active) up until roughly 60% of the yield strain. As strain continues to increase, the rejuvenation mechanism becomes active and $S_0$ begins to gradually increase, experiencing the fastest increases early in the strain softening regime. Saturation of $S_0$ (and the rejuvenation mechanism) does not occur until strains at least eight times the yield strain. In Figure 5, we have used calculations provided by the theory for a PMMA glass with conditions similar to those of our experiments ($T_g$ - 10 K, $10^{-3}$ $s^{-1}$ constant strain rate, pre-aging time = $10^5$ s); theoretical calculations for different temperatures, strain rates, and aging times demonstrate very similar trends in rejuvenation activity.[9] Calculations in the framework of the NLE theory require several material-dependent input parameters as described in an earlier report;[10] for PMMA, these were obtained from the experimental density, temperature-dependent compressibility and x-ray scattering measurements, and by requiring the theory to reproduce $T_g$ and $T_c$ (a characteristic high temperature crossover temperature for dynamics). In addition, the calculations that we reproduce from reference 9 utilized a statistical segment length of 1 nm and one additional parameter that fixes the extent to which plastic work results in rejuvenation.[9]

Figure 5 shows good general agreement in the post-yield regime between the NLE theory and the calculations of rejuvenation based upon the two sets of experimental measurements presented here. In all cases, rejuvenation gradually develops as a function of strain up roughly five times the yield strain. However, in the pre-yield regime, the probe reorientation measurements show systematically higher levels of rejuvenation than both the mechanical measurements and the theory.

We emphasize that the NLE calculation reproduced in Figure 5 reports $S_0$ *during constant strain rate deformation* and not after the reversing strain that was employed in the experiment. For this reason, the comparison in Figure 5 is not completely quantitative. Clearly it would be useful to compare the experimental results to NLE calculations that mimic the full experimental protocol; an example of such calculations are presented in very recent work.[49] However, we emphasize that the discrepancy between the NLE theory and the probe reorientation experiments in the pre-yield regime will not be resolved by including the reversing strain in the calculation; if $S_0$ does not increase during the forward deformation, the structure of the theory does not allow it to increase as the stress returns to zero (see equation 10 in reference 9). We remind the reader that the reversing strain is an essential part of the experimental comparison with the NLE approach since only in the zero-stress condition can we rigorously separate the contribution of rejuvenation from stress activation.

We speculate that the qualitative difference in Figure 5 between the probe reorientation method and the NLE calculation for pre-yield strains results from spatially heterogeneous dynamics. The NLE theory is a single relaxation time approach, while it is known that supercooled liquids and glasses display spatially heterogeneous dynamics.[50-53] Furthermore, it is known that the distribution of relaxation times can change during deformation.[20, 26, 27, 30, 31] In a polymer glass with spatially heterogeneous dynamics, it is natural to imagine that local regions do not all yield at the same strain. This is consistent with computer simulations indicating that plastic deformation occurs first in soft spots that act as structural “defects” in the glassy structure.[53, 54] Those local regions that yield first presumably also undergo local rejuvenation that results in segmental dynamics that remain enhanced even after stress is removed by the reversing deformation. From this perspective, it is qualitatively reasonable that rejuvenation begins at lower strains in a spatially heterogeneous system than is predicted by the mean-field NLE theory. The Stochastic Constitutive Model (SCM) of Medvedev and Caruthers[6] provides a useful insight on this point because of its ability to track how the distribution of relaxation times changes during aging and deformation. According to the SCM model,[6] during a constant strain rate deformation, the relaxation spectrum drastically narrows before yield, in qualitative agreement with experimental results.[48] It would be useful to more thoroughly characterize the heterogeneous dynamics of a deforming polymer glass by following the dynamics of individual probe molecules; recent experimental advances indicate that this experiment is feasible. (ACS Nano 2016, 10, 1908−1917).

We remark that the experimental results presented here should be reproduced by any theory which correctly describes changes in segmental dynamics during deformation, including those using an framework different than the NLE theory. For example, it would be interesting to see if the SCM model can reproduce these results, even though this model does not naturally decompose segmental mobility into contributions from landscape tilting and mechanical rejuvenation. Such a result would provide independent evidence for the validity of the SCM approach and might additionally indicate that these two concepts are relevant for understanding deformation-enhanced segmental dynamics within this model.

*Comparison of probe reorientation results with previous computer simulations*

Smessaert and Rottler[44] have simulated the segmental dynamics of a polymer glass during the recovery period after reversing constant stress and constant strain rate deformations. For these simulations, they utilized a generic bead-spring model. The segmental relaxation time $\tau$ in the simulation represents a characteristic time for the decay of the intermediate scattering function.

We compare our probe reorientation times following the reversing deformation to selected results from the constant strain rate simulations in Figure 6. The format of Figure 6 is taken from reference 44. The y-axis shows $\tau$ (simulation) or $\tau_{1/e}$ (experiment, from Figure 3), normalized to its value in the absence of deformation at an aging time of $t_a$ (the total elapsed time immediately after the reversing deformation). The x-axis normalizes $t_r$, the time elapsed in the recovery period after stress is removed, to $t_a$. The strains in the legend reflect the maximum attained strain in the glass (simulation) or local measurement area (experiment) during the deformation; these strains are normalized to the appropriate yield strain (2.9% in the experiment, 6.0% in the simulation). For ease of comparison with the experiments, the simulation data have been smoothed and are presented by continuous curves. Because the aging exponent μ was somewhat different in the experiment and simulation ($\mu$=0.72 in the experiment, $\mu$=0.89 in the simulation),[44] we have slightly adjusted the simulation results in order to facilitate a comparison with the experiments; the simulation log τ values were multiplied by a factor of 0.72/0.89 to account for the difference of the μ exponents.

As explained in reference 44, the format of Figure 6 is motivated by the idea that deformed glasses in many cases will be governed by their predeformation aging behavior. For a glass with no or very little imposed deformation, physical aging continues without interruption. For this pure aging case, the dynamics approximately follow a horizontal line at $\tau_{1/e} / \tau_{1/e,0} = 1$ until a value of $t_r/t_a = 1$ is approached, after which the continuous aging of the glass is observed to follow the characteristic slope μ associated with the generic aging behavior (solid black line in Figure 6). For highly deformed glasses that have been rejuvenated (as evidenced by their small τ values), aging in the simulation begins immediately after deformation and the dynamics in the simulation roughly follow the generic aging line even for times $t_r/t_a < 1$.

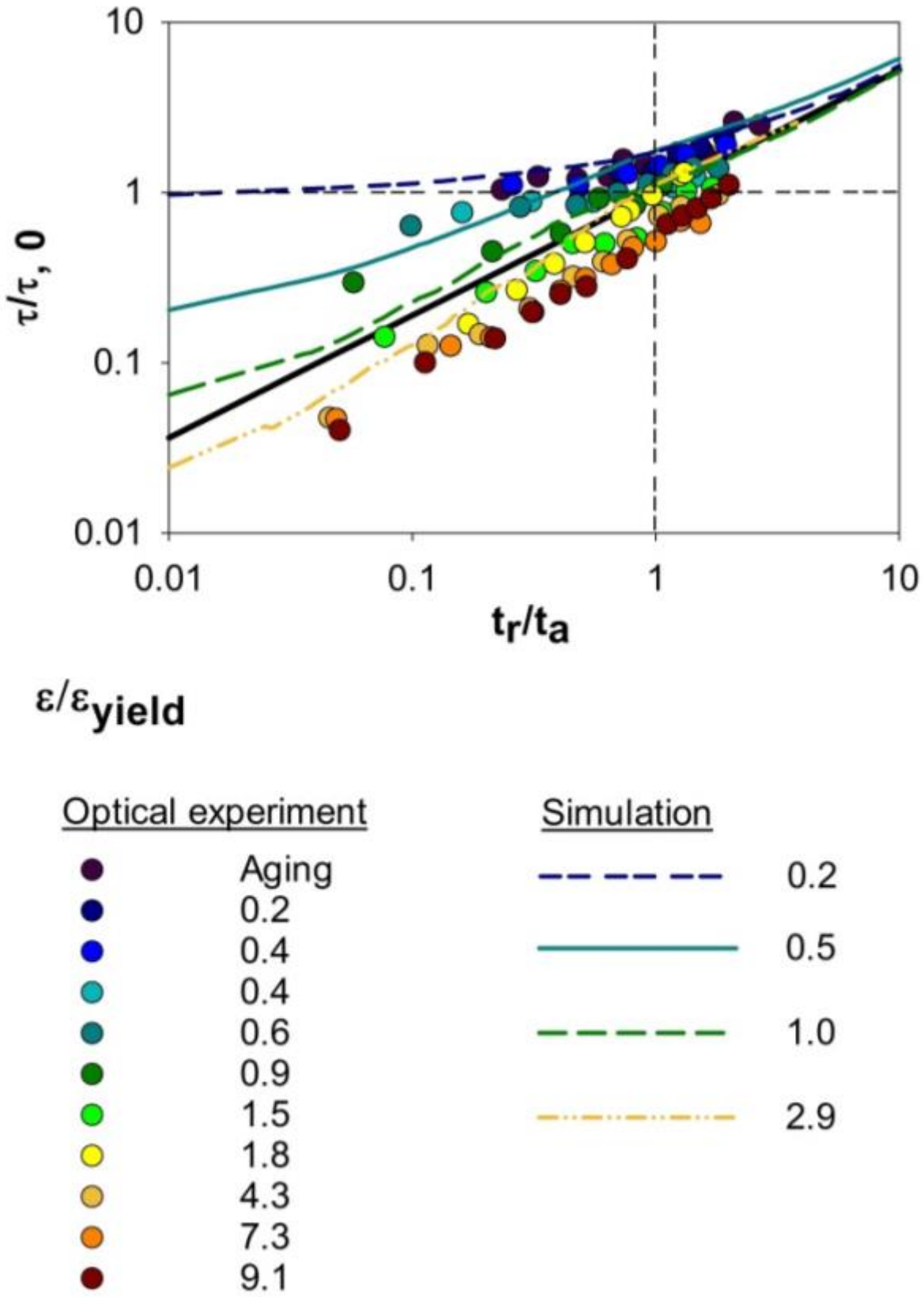

**Figure 6.** Post-deformation τ values plotted against recovery time after reversing deformations, for both the present work (data points) and the simulations of reference 44 (lines). The y-axis shows $\tau_{1/e}$ (present work) or τ (simulation), normalized to its quiescent value at the end of deformation; the x-axis normalizes the elapsed time in the recovery regime to the total aging time of the sample at the end of the deformation. The solid black line indicates slope μ associated with the generic aging behavior in the experiments (μ = 0.72). The simulation results have been smoothed and adjusted for differing μ as described in the text. Values in the legend represent the maximum strain attained during deformation, normalized to the yield strain.

As can be seen in Figure 6, the simulations of Smessaert and Rottler and our experiments show strikingly similar initial enhancements in dynamics (immediately following the reversing deformation) and also similar recovery behavior. The simulation work supports our findings of rejuvenation activity (as indicated by enhanced segmental dynamics) for pre-yield reversing deformations. For example, simulation results for a strain that reached 50% of yield show significant departure from the aging trajectory (see solid teal curve). Simulation results at additional strains reported by Smessaert and Rottler[44] (not shown) are also consistent with a gradual increase in enhanced dynamics, even when very low strains were reached in the reversing deformation. The agreement of the low-strain simulation and optical results of Figure 6 supports the view that substantial rejuvenation occurs before yield (even if this does not result in a significant impact on the magnitude of the stress overshoot, as we discuss below). As strain increases in the simulations of Smessaert and Rottler,[44] the enhancement of dynamics that results from the reversing deformation increases and then saturates at about three times the yield strain. This feature of the simulations is also qualitatively consistent with the optical experiments reported in this work as shown in Figures 5 and 6.

Although the simulation[44] and optical results broadly agree, at large strains the experimental results do not recover the generic aging behavior of the glass (solid black line) at a recovery time equal to the original age of the glass at the end of deformation ($t_r/t_a = 1$). Rather, the optical measurements show that, for the largest deformations, dynamics are still significantly enhanced (relative to the non-deformed glass) at these long times. This enhancement of dynamics would seem to indicate that the PMMA glass is aging on a different landscape after deformation, perhaps as a result of structural changes (e.g., anisotropy or a change in density). Consistent with this view, Rottler and Smessaert reported changes in structural variables after very large strains in their simulations and concluded that the deformed glass represented a distinct thermodynamic state. These observations seem consistent with arguments by McKenna[35] that polymer glass deformation can be described as a polyamorphic phase transition.

We point out that slight differences exist in the mechanical protocols used in the simulations of reference 44 and in our experiments; the experiment applies and reverses deformation at the same rate, while in the simulations, stress is unloaded at a near-instantaneous rate. However, because the unloading period in the experiment represents a small fraction of the pre-deformation aging time (1.1% at maximum), this difference in protocol is unlikely to affect the qualitative comparison between simulation and experiment at long times.

Lyulin and coworkers[55] investigated signatures of mechanical rejuvenation in a simulated atactic polystyrene glass during and after large cyclic shear deformations. In these simulations the observed storage and loss moduli evolved towards steady state values during several shear cycles. Lyulin and coworkers[55] extracted a measure of segmental dynamics ($\tau$) by monitoring the reorientation of a specific bond vector within their atomistic model. The behavior of their extracted $\tau$ values before and after cyclic deformation are qualitatively consistent with the reported results of Figure 3, with decreased $\tau$ values immediately following deformation, followed by lengthening $\tau$ values during recovery. We note that the work of Lyulin and coworkers[55] was performed on capped thin films and demonstrates confinement effects; however we anticipate that the qualitative features associated with rejuvenation are not altered by confinement.

Lee et al. previously reported measurements of segmental dynamics for a protocol in which creep deformation (e.g., constant stress) was followed by aging at zero stress.[29] In that work, both pre-flow and flow-state constant stress deformations were applied. After pre-flow deformation, segmental dynamics were transiently faster with the original aging behavior recovered within an additional time period of $t_a$.[29] This behavior is consistent with the pre-yield results reported in Figure 6. For deformations which reached the flow state, Lee et al. reported two behaviors, only one of which matches the results presented here. For low temperature deformations, they reported (see Figure 12 of reference 29) that segmental dynamics following deformation remained significantly faster than expected given the original aging trajectory even for times greater than $t_a$; this is similar to the results of the experiments shown in Figure 6. On the other hand, for high deformation temperatures (including the same temperature used in this work), they reported that the aging behavior following a large deformation matched the original aging behavior after a time shift similar to the original aging time; this behavior does not match the present experiments but is similar to the results reported in the simulations of Rottler and Smessaert.[44] We are unsure how to interpret the differences between the present results and those reported by Lee et al. The most obvious difference in the experiments is the deformation protocol: constant strain rate extension and retraction here vs. creep extension and stress-free recovery in reference 29. Further experiments comparing a variety of protocols may be required to understand the post-deformation aging behavior.

*Comparisons among different measures of rejuvenation*

As described in the introduction, there has been controversy about mechanical rejuvenation, involving both its defining characteristics and whether it is activated by pre-yield deformation. In this section, we compare our new measurements with experimental results in the literature, particularly with regard to these controversial points.

We first compare our own probe reorientation measurements with our stress overshoot measurements. As shown in Figure 5, while there is broad agreement for these two observables with respect to the way in which rejuvenation is activated with strain, there is disagreement in the pre-yield regime. Since the pre-yield deformations are macroscopically homogeneous,[30, 31, 48] this result cannot be attributed to differences between global and local strains. We note that our stress overshoot measurements are in qualitative agreement with previous stress overshoot measurements and thus serve as a validation of the mechanical aspects of our experiments. In our view, it is difficult to determine the extent of rejuvenation from the stress overshoot experiment. In order to test the state of the sample after the initial reversing deformation, a highly nonlinear perturbation (constant strain-rate deformation through yield) is performed. One can imagine that deformation through yield would rejuvenate all samples to a similar state, independent of whether any pre-yield rejuvenation had occurred earlier. Given this argument (for which we thank an anonymous reviewer) and previous indications that even small mechanical perturbations are difficult to interpret if they follow a nonlinear deformation,[56] we conclude that the stress overshoot measurement cannot be simply interpreted in terms of rejuvenation. Also, we note that the relationship between the optical and mechanical experiments might depend upon the deformation temperature and strain rate. In addition, recent simulations have shown that chain orientation can also have an important impact on the magnitude of the stress overshoot in pre-deformed glasses,[57] and this also can obscure the connection with rejuvenation. In spite of these complications, the stress overshoot is an important mechanical property; an accurate model of polymer glass deformation should be able to describe the behavior of both experimental observables shown in Figure 5, including their differences.

We next compare our probe reorientation measurements to the simultaneous deformation and volume measurements of McKenna and coworkers.[35, 41] In our view, this experiment is conceptually similar to our simultaneous deformation and probe reorientation measurements. In both experiments, an aging polymer glass is subjected to a reversing nonlinear deformation (torsion in reference 41 and extension/contraction here) followed by a “read out” of the state of the sample that does not perturb the state of system (volume in reference 41 and probe reorientation time here). A further similarity is that both the volume and the probe reorientation time can be measured continuously after the reversing deformation in order to determine the time response of the system to the reversing deformation. Although the materials used in these experiments were different (epoxy[41] vs cross-linked PMMA), the temperatures relative to Tg were similar ($T_g$ – 10 K[41] vs. $T_g$ – 7 K).

While McKenna and coworkers[41] observed an instantaneous change in the sample volume associated with the twisting and untwisting of their sample, the most interesting feature for our discussion is the way in which the perturbed volume returns to the volume observed for an unperturbed glass. They reported that the volume response to the perturbation died away much more quickly (~ 50 times faster) than the overall aging time and that the underlying aging trajectory of the glass was not altered even by a reversing torsional strain of 5% (near yield). In contrast, the probe reorientation data shown in Figures 3 and 6 shows a slow recovery back to the unperturbed aging curve. In the pre-yield regime, perturbations that last less than 200 s give rise to effects that can still be observed 20000 s after the perturbation. Furthermore, the analysis shown in Figure 6 is consistent with the idea that the aging clock was reset by the pre-yield deformation (although we did not measure the time to reach equilibrium and so we cannot compare to this aspect of the work in reference 41). Of course, the differences between our work and that of reference 41 might also be responsible for the apparent differences in the results (different materials, different deformation, and different observables used to determine deviation from equilibrium), and this should be considered in the interpretation of these results.

The differential scanning calorimetry (DSC) work of Hasan and Boyce also provide an important point of comparison with our results.[43] They performed DSC measurements on a number of polymer glasses after constant strain rate deformations which were unloaded at varying strains. While there are many interesting aspects of their results, the most pertinent observation for our discussion is the change in enthalpy between an annealed polymer glass and one that had been deformed in the pre-yield regime. Although they do not extensively sample the pre-yield regime, Hasan and Boyce[43] report only a small change in enthalpy in comparison to that observed for much larger strains. In this respect, the enthalpy behaves much more like the NLE calculation shown in Figure 5, i.e., the qualitative interpretation of their results is that rejuvenation is a small effect in the pre-yield regime. As in the case of the comparison with the work of McKenna and coworkers,[41] we emphasize that different observables are not required to respond identically to reversing deformations. The enthalpy measured by the DSC experiments provides an indication of the depths of the potential minima for the system in the PEL while the optical measurements of the dynamics are sensitive to the heights of barriers.

An accurate model of polymer glass deformation should be able to reproduce all of these experimental results. While the structure of the NLE model motivated the design of these experiments (since one of two mechanisms in that approach can be turned off by setting the stress to zero), it is possible that a theory with a completely different structure might be able to reproduce the probe reorientation and mechanical data sets, along with other studies such as those discussed above.

**Concluding Remarks**

In this work, we have investigated the mechanisms that control enhanced segmental dynamics during the deformation of polymer glasses. We compared the evolution of two different measures of rejuvenation as function of strain: a purely mechanical measurement based upon yield stress and an optical measurement of segmental dynamics. Both measurements indicate that rejuvenation gradually develops and saturates at strains at least five times the yield strain. These experimental results are broadly consistent with the NLE theory of Chen and Schweizer.[8, 9] In this respect, the results support the view that enhanced segmental dynamics during the deformation of polymer glasses results from both rejuvenation (pulling the system up the PEL) and stress activation (tilting the PEL).

For deformations which are macroscopically pre-yield, significantly larger levels of rejuvenation are indicated by the segmental dynamics than by the mechanical measure of rejuvenation or the NLE theory. Our finding of enhanced segmental dynamics after reversing pre-yield deformations is consistent with the simulations of Smessaert and Rottler.[44] We interpret the pre-yield rejuvenation to be a result of spatially heterogeneous dynamics, an effect not included in the NLE theory.[8, 9] Our mechanical results highlight a limit of using the yield stress as an indicator of rejuvenation since probing rejuvenation in this manner is a highly nonlinear process.

One of the most interesting comparisons possible with this new work is that with the combined mechanical/volume measurements of McKenna and coworkers.[35, 41] Those measurements indicate that pre-yield deformation does not perturb the underlying aging trajectory of a polymer glass. In contrast, our probe reorientation results indicate the opposite conclusion. Since two different observables are utilized, this is not necessarily a contradiction and it is possible that both behaviors are exhibited simultaneously. Future efforts should work to integrate these two perspectives.

**Acknowledgements**

We thank the National Science Foundation (DMR – 1404614) for support of this research. We thank Josh Ricci, Kelly Suralik, Trevor Bennin, Lian Yu, and Travis Powell for assistance with experiments and helpful discussions. We additionally thank Ken Schweizer, Joerg Rottler, Jim Caruthers, Grigori Medvedev, Anton Smessaert, and Michael Tylinski for helpful discussions.

## References

1. Eyring, H. Viscosity, Plasticity, and Diffusion as Examples of Absolute Reaction Rates. *Journal of Chemical Physics* **1936,** *4*, 283-291.

2. Klompen, E. T. J.; Engels, T. A. P.; Govaert, L. E.; Meijer, H. E. H. Modeling of the Postyield Response of Glassy Polymers: Influence of Thermomechanical History. *Macromolecules* **2005,** *38*, 6997-7008.

3. De Focatiis, D. S. A.; Embery, J.; Buckley, C. P. Large deformations in oriented polymer glasses: Experimental study and a new glass-melt constitutive model. *Journal of Polymer Science Part B: Polymer Physics* **2010,** *48*, 1449-1463.

4. Schapery, R. A. An engineering theory of nonlinear viscoelasticity with applications. *International Journal of Solids and Structures* **1966,** *2*, 407-425.

5. Fielding, S. M.; Larson, R. G.; Cates, M. E. Simple Model for the Deformation-Induced Relaxation of Glassy Polymers. *Physical Review Letters* **2012,** *108*, 048301.

6. Medvedev, G. A.; Caruthers, J. M. Development of a stochastic constitutive model for prediction of postyield softening in glassy polymers. *Journal of Rheology* **2013,** *57*, 949-1002.

7. Knauss, W. G.; Emri, I. Volume change and the nonlinearly thermo-viscoelastic constitution of polymers. *Polymer Engineering & Science* **1987,** *27*, 86-100.

8. Chen, K.; Schweizer, K. S. Theory of aging, rejuvenation, and the nonequilibrium steady state in deformed polymer glasses. *Physical Review E* **2010,** *82*, 041804.

9. Chen, K.; Schweizer, K. S. Theory of Yielding, Strain Softening, and Steady Plastic Flow in Polymer Glasses under Constant Strain Rate Deformation. *Macromolecules* **2011,** *44*, 3988-4000.

10. Chen, K.; Schweizer, K. S. Microscopic Constitutive Equation Theory for the Nonlinear Mechanical Response of Polymer Glasses. *Macromolecules* **2008,** *41*, 5908-5918.

11. Chen, K.; Schweizer, K. S. Theory of relaxation and elasticity in polymer glasses. *Journal of Chemical Physics* **2007,** *126*, 014904.

12. Meijer, H. E. H.; Govaert, L. E. Mechanical performance of polymer systems: The relation between structure and properties. *Progress in Polymer Science* **2005,** *30*, 915-938.

13. Caruthers, J. M.; Adolf, D. B.; Chambers, R. S.; Shrikhande, P. A thermodynamically consistent, nonlinear viscoelastic approach for modeling glassy polymers. *Polymer* **2004,** *45*, 4577-4597.

14. Pschorn, U.; Roessler, E.; Sillescu, H.; Kaufmann, S.; Schaefer, D.; Spiess, H. W. Local and cooperative motions at the glass transition of polystyrene: information from one- and two-dimensional NMR as compared with other techniques. *Macromolecules* **1991,** *24*, 398-402.

15. Riggleman, R. A.; Toepperwein, G. N.; Papakonstantopoulos, G. J.; de Pablo, J. J. Dynamics of a Glassy Polymer Nanocomposite during Active Deformation. *Macromolecules* **2009,** *42*, 3632-3640.

16. Warren, M.; Rottler, J. Deformation-induced accelerated dynamics in polymer glasses. *Journal of Chemical Physics* **2010,** *133*, 164513.

17. Chung, Y. G.; Lacks, D. J. How Deformation Enhances Mobility in a Polymer Glass. *Macromolecules* **2012,** *45*, 4416-4421.

18. Chung, Y. G.; Lacks, D. J. Atomic mobility in strained glassy polymers: The role of fold catastrophes on the potential energy surface. *Journal of Polymer Science Part B: Polymer Physics* **2012,** *50*, 1733-1739.

19. Lee, H. N.; Riggleman, R. A.; de Pablo, J. J.; Ediger, M. D. Deformation-Induced Mobility in Polymer Glasses during Multistep Creep Experiments and Simulations. *Macromolecules* **2009,** *42*, 4328-4336.

20. Riggleman, R. A.; Lee, H. N.; Ediger, M. D.; de Pablo, J. J. Heterogeneous dynamics during deformation of a polymer glass. *Soft Matter* **2010,** *6*, 287-291.

21. Loo, L. S.; Cohen, R. E.; Gleason, K. K. Chain mobility in the amorphous region of nylon 6 observed under active uniaxial deformation. *Science* **2000,** *288*, 116-119.

22. Zhou, Q. Y.; Argon, A. S.; Cohen, R. E. Enhanced Case-II diffusion of diluents into glassy polymers undergoing plastic flow. *Polymer* **2001,** *42*, 613-621.

23. Kalfus, J.; Detwiler, A.; Lesser, A. J. Probing Segmental Dynamics of Polymer Glasses during Tensile Deformation with Dielectric Spectroscopy. *Macromolecules* **2012,** *45*, 4839-4847.

24. Pérez-Aparicio, R.; Cottinet, D.; Crauste-Thibierge, C.; Vanel, L.; Sotta, P.; Delannoy, J.-Y.; Long, D. R.; Ciliberto, S. Dielectric Spectroscopy of a Stretched Polymer Glass: Heterogeneous Dynamics and Plasticity. *Macromolecules* **2016,** *49*, 3889-3898.

25. Lee, H. N.; Paeng, K.; Swallen, S. F.; Ediger, M. D. Dye reorientation as a probe of stress-induced mobility in polymer glasses. *Journal of Chemical Physics* **2008,** *128*, 134902.

26. Lee, H. N.; Paeng, K.; Swallen, S. F.; Ediger, M. D. Direct Measurement of Molecular Mobility in Actively Deformed Polymer Glasses. *Science* **2009,** *323*, 231-234.

27. Lee, H. N.; Paeng, K.; Swallen, S. F.; Ediger, M. D.; Stamm, R. A.; Medvedev, G. A.; Caruthers, J. M. Molecular Mobility of Poly(methyl methacrylate) Glass During Uniaxial Tensile Creep Deformation. *Journal of Polymer Science Part B: Polymer Physics* **2009,** *47*, 1713-1727.

28. Lee, H. N.; Ediger, M. D. Interaction between physical aging, deformation, and segmental mobility in poly(methyl methacrylate) glasses. *Journal of Chemical Physics* **2010,** *133*, 014901.

29. Lee, H. N.; Ediger, M. D. Mechanical Rejuvenation in Poly(methyl methacrylate) Glasses? Molecular Mobility after Deformation. *Macromolecules* **2010,** *43*, 5863-5873.

30. Bending, B.; Christison, K.; Ricci, J.; Ediger, M. D. Measurement of Segmental Mobility during Constant Strain Rate Deformation of a Poly(methyl methacrylate) Glass. *Macromolecules* **2014,** *47*, 800-806.

31. Hebert, K.; Bending, B.; Ricci, J.; Ediger, M. D. Effect of Temperature on Postyield Segmental Dynamics of Poly(methyl methacrylate) Glasses: Thermally Activated Transitions Are Important. *Macromolecules* **2015,** *48*, 6736-6744.

32. Ediger, M. D.; Hebert, K., A molecular perspective on the yield and flow of polymer glasses: The role of enhanced segmental dynamics during active deformation. In *Polymer Glasses*, Roth, C. B., Ed. Taylor & Francis: Submitted.

33. Govaert, L. E.; van Melick, H. G. H.; Meijer, H. E. H. Temporary toughening of polystyrene through mechanical pre-conditioning. *Polymer* **2001,** *42*, 1271-1274.

34. van Melick, H. G. H.; Govaert, L. E.; Raas, B.; Nauta, W. J.; Meijer, H. E. H. Kinetics of ageing and re-embrittlement of mechanically rejuvenated polystyrene. *Polymer* **2003,** *44*, 1171-1179.

35. McKenna, G. B. Mechanical rejuvenation in polymer glasses: fact or fallacy? *Journal of Physics: Condensed Matter* **2003,** *15*, S737.

36. Bauwens-Crowet, C.; Bauwens, J. C. Effect of annealing on the shear yield stress of rejuvenated polycarbonate. *Polymer* **1988,** *29*, 1985-1989.

37. McKenna, G. B.; Kovacs, A. J. Physical aging of poly(methyl methacrylate) in the nonlinear range: Torque and normal force measurements. *Polymer Engineering & Science* **1984,** *24*, 1138-1141.

38. Debenedetti, P. G.; Stillinger, F. H. Supercooled liquids and the glass transition. *Nature* **2001,** *410*, 259-267.

39. Riggleman, R. A.; Schweizer, K. S.; de Pablo, J. J. Nonlinear Creep in a Polymer Glass. *Macromolecules* **2008,** *41*, 4969-4977.

40. Struik, L. C. E., *Physical Aging in Amorphous Polymers and Other Materials*. Elsevier: Amsterdam, 1978.

41. Santore, M. M.; Duran, R. S.; McKenna, G. B. Volume recovery in epoxy glasses subjected to torsional deformations: the question of rejuvenation. *Polymer* **1991,** *32*, 2377-2381.

42. Struik, L. C. E. On the rejuvenation of physically aged polymers by mechanical deformation. *Polymer* **1997,** *38*, 4053-4057.

43. Hasan, O. A.; Boyce, M. C. Energy storage during inelastic deformation of glassy polymers. *Polymer* **1993,** *34*, 5085-5092.

44. Smessaert, A.; Rottler, J. Recovery of Polymer Glasses from Mechanical Perturbation. *Macromolecules* **2012,** *45*, 2928-2935.

45. D1708-10, Standard Test Method for Tensile Properties of Plastics by Use of Microtensile Specimens. ASTM International: West Conshohocken, PA, 2011; pp 1-5.

46. Lee, E. W.; Medvedev, G. A.; Caruthers, J. M. Deformation induced evolution of mobility in PMMA. *Journal of Polymer Science Part B: Polymer Physics* **2010,** *48*, 2399-2401.

47. Maher, J. W.; Haward, R. N.; Hay, J. N. Study of the thermal effects in the necking of polymers with the use of an infrared camera. *Journal of Polymer Science: Polymer Physics Edition* **1980,** *18*, 2169-2179.

48. Bending, B.; Ediger, M. D. Comparison of mechanical and molecular measures of mobility during constant strain rate deformation of a PMMA glass. *Journal of Polymer Science Part B: Polymer Physics* **2016,** *54*, 1957-1967.

49. Rottler, J. Relaxation times in deformed polymer glasses: A comparison between molecular simulations and two theories. *The Journal of Chemical Physics* **2016,** *145*, 064505.

50. Ediger, M. D. Spatially heterogeneous dynamics in supercooled liquids. *Annual Review of Physical Chemistry* **2000,** *51*, 99-128.

51. Richert, R. Heterogeneous dynamics in liquids: fluctuations in space and time. *Journal of Physics-Condensed Matter* **2002,** *14*, R703-R738.

52. Ediger, M. D.; Harrowell, P. Perspective: Supercooled liquids and glasses. *Journal of Chemical Physics* **2012,** *137*, 080901.

53. Smessaert, A.; Rottler, J. Structural relaxation in glassy polymers predicted by soft modes: a quantitative analysis. *Soft Matter* **2014,** *10*, 8533-8541.

54. Berthier, L.; Ediger, M. D. Facets of glass physics. *Physics Today* **2016,** *69*, 40-46.

55. Hudzinskyy, D.; Michels, M. A. J.; Lyulin, A. V. Rejuvenation, Aging, and Confinement Effects in Atactic-Polystyrene Films Subjected to Oscillatory Shear. *Macromolecular Theory and Simulations* **2013,** *22*, 71-84.

56. McKenna, G. B.; Zapas, L. J. Superposition of small strains on large deformations as a probe of nonlinear response in polymers. *Polymer Engineering & Science* **1986,** *26*, 725-729.

57. Liu, A. Y. H.; Rottler, J. Aging under stress in polymer glasses. *Soft Matter* **2010,** *6*, 4858-4862.

**For Table of Contents Use Only**

**Reversing strain deformations probe mechanisms for enhanced segmental mobility of polymer glasses**

Kelly Hebert and M.D. Ediger

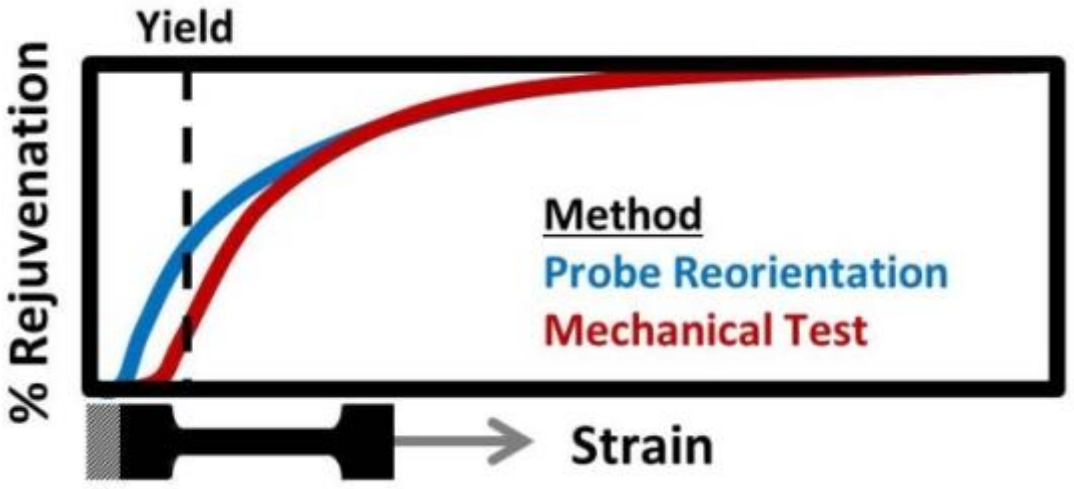